\PassOptionsToPackage{table}{xcolor}
\documentclass[10pt, a4paper, copyright]{krafton-ai}

\unless\ifpdf\microtypesetup{tracking=false}\fi

\usepackage[authoryear, sort&compress, round]{natbib}
\usepackage{wrapfig}
\usepackage[table]{xcolor}
\usepackage{booktabs}  %
\usepackage{tabularx}  %
\usepackage{ragged2e}  %
\usepackage{makecell}
\usepackage{xspace}
\usepackage{pifont}
\usepackage{subcaption}
\usepackage{graphicx}
\usepackage{multirow}
\usepackage[most]{tcolorbox}
\tcbuselibrary{listings,breakable}

\lstdefinestyle{bt}{
  basicstyle=\ttfamily\small,
  backgroundcolor=\color{black!6},
  frame=single,
  rulecolor=\color{black!30},
  frameround=tttt,
  framesep=6pt,
  numbers=left,
  numberstyle=\tiny\color{black!50},
  numbersep=10pt,
  xleftmargin=2.2em,
  showstringspaces=false,
  columns=fullflexible,
  keepspaces=true,
  keywordstyle=\bfseries\color{teal!70!black},
  morekeywords={selector,sequence,condition,task},
}

\NewTColorBox[auto counter]{example}{ O{!htbp} m O{} }{
  enhanced,
  breakable,
  float*,
  floatplacement={#1},
  width=\textwidth,
  colback=gray!5!white,
  colframe=gray!75!black,
  title={Example~\thetcbcounter: #2},
  enlarge top by=5mm,
  enlarge bottom by=5mm,
  label={#3}
}

\newif\ifreview
\reviewtrue   %

\ifreview
  
\else
  
\fi

\newcommand{\RaonPool}{Raon-OpenTTS-Pool\xspace}
\newcommand{\RaonCore}{Raon-OpenTTS-Core\xspace}
\newcommand{\RaonModel}{Raon-OpenTTS\xspace}
\newcommand{\RaonEval}{Raon-OpenTTS-Eval\xspace}
\newcommand{\RaonYouTube}{Raon-YouTube-Commons\xspace}

\newcommand{\cmark}{{\color{green!60!black}\ding{51}}}

\uselogo{} 

\title{Raon-OpenTTS: Open Models and Data for Robust Text-to-Speech}

\usepackage[useregional=false]{datetime2}
\DTMsetstyle{iso}
\paperdate{\DTMtoday} %

\author[3]{Semin Kim$^{*\dagger}$}
\author[1]{Seungjun Chung$^{*}$}
\author[1]{Taehong Moon}
\author[1,4]{Sangheon Lee$^{\dagger}$}
\author[1,3]{Minyoung Ahn$^{\dagger}$}
\author[1]{Keon Lee}
\author[3]{Nam~Soo Kim}
\author[1]{Jaewoong Cho}
\author[6]{Ludwig Schmidt}
\author[1,2,5]{Kangwook Lee}
\author[1]{Dongmin Park}

\affil[1]{KRAFTON}
\affil[2]{Ludo Robotics}
\affil[3]{Seoul National University}
\affil[4]{KAIST}
\affil[5]{University of Wisconsin-Madison}
\affil[6]{Stanford University}

\begin{abstract}
  Recent advances in text-to-speech~(TTS) models show impressive speech naturalness and quality,
  yet the role of large-scale open data in driving this progress remains underexplored.
  In this work, we introduce \textbf{\RaonModel}, an open TTS model that performs competitively with state-of-the-art closed-data TTS models,
  and \textbf{\RaonPool}, a large-scale open dataset for reproducible TTS training.
  \RaonPool consists of \textbf{615K hours} of \textbf{240M} speech segments aggregated from publicly available English speech corpora and web-sourced recordings.
  With a model-based filtering pipeline applied to \RaonPool, we derive \textbf{\RaonCore},
  a curated, high-quality subset of 510K hours and 194M speech segments.
  Using \RaonCore, we train \RaonModel, a series of diffusion transformer~(DiT)-based TTS models from 0.3B to 1B parameters. 
  On multiple benchmarks, \RaonModel-1B shows comparable performance to state-of-the-art models such as Qwen3-TTS and CosyVoice~3, which are trained on several million hours of proprietary speech data.
  Notably, on Seed-TTS-Eval, \RaonModel-1B achieves a word error rate~(WER) of 1.78$\%$ and a speaker similarity~(SIM) of 0.749, ranking second on WER and first on SIM among recent open-weight TTS baselines.
  On CV3-Hard-EN, \RaonModel-1B achieves a WER of 6.15$\%$ and a SIM of 0.775,
  ranking first on both metrics.
  Furthermore, to support robust evaluation, we introduce \textbf{\RaonEval},
  a structured benchmark for assessing TTS robustness across diverse acoustic conditions
  including clean, noisy, in-the-wild, and expressive speech.
  On \RaonEval, \RaonModel-1B achieves the best average WER and SIM
  among all evaluated models, and the second-best human preference, as measured by comparative mean opinion score~(CMOS).
  Our data pool, filtering pipeline, training code, and checkpoints are publicly available
  at \url{https://github.com/krafton-ai/RAON-OpenTTS}.
\end{abstract}

\begin{document}

\maketitle
{\let\thefootnote\relax\footnotetext{$^*$ Equal contribution.}}
{\let\thefootnote\relax\footnotetext{$^\dagger$ Work done while doing an internship at KRAFTON.}}

\section{Introduction}
\label{sec:intro}
Text-to-speech (TTS) synthesis has advanced rapidly in recent years, with modern models capable of producing high-quality speech across diverse speakers, styles, and acoustic conditions. Despite the growing number of TTS models, most state-of-the-art models do not fully disclose their training data or data curation pipelines, which limits the research community to systematically study, reproduce, and build upon these advances. Just as open data efforts have played a critical role in accelerating progress in other domains, such as those underlying large language models (LLMs)~\citep{penedo2024fineweb}, CLIP~\citep{gadre2023datacomp}, and Whisper~\citep{ngo2025olmoasr}, establishing open datasets and models for TTS is essential to enable systematic and reproducible research at scale.

State-of-the-art TTS models trained on large-scale proprietary data, such as Qwen3-TTS~\citep{hu2026qwen3} and CosyVoice~3~\citep{du2025cosyvoice}, achieve remarkable zero-shot speech generation performance across diverse speakers and acoustic conditions. However, these models are trained on multi-million hours of proprietary speech data unavailable to the public. Among models trained entirely on open speech-text data, F5-TTS~\citep{chen2025f5} and MaskGCT~\citep{wang2024maskgct} show the most competitive performance. Both models are trained on approximately 100K hours of a single speech dataset, Emilia~\citep{he2024emilia}. Nevertheless, as shown in Table~\ref{tab:seed-eval}, their performance still exhibits a significant gap compared to state-of-the-art systems trained on proprietary data, highlighting the need for a comprehensive study on whether large-scale, multi-source open datasets can bridge this gap and enable fully reproducible models.

In this paper, we establish a high-performing, fully reproducible TTS framework with open data, open weights, and a transparent training pipeline. We first present \RaonPool, a 615K-hour English speech-text dataset constructed from 10 publicly available speech corpora together with web-sourced recordings. From this pool, we derive \RaonCore, a filtered training subset constructed using a model-based pipeline that removes potentially problematic speech segments based on transcription consistency, acoustic quality, and speech activity ratio. Next, we release \RaonModel, a series of diffusion transformer (DiT)-based TTS models with 0.3B and 1B parameters, trained on our proposed datasets. 
Notably, \RaonModel-1B trained on \RaonCore achieves performance comparable to state-of-the-art TTS models across popular TTS benchmarks such as Seed-TTS-Eval~\citep{anastassiou2024seed} and CV3-Eval~\citep{du2025cosyvoice}, ranking first or second in word error rate (WER) and speaker similarity (SIM) among recent zero-shot TTS baselines.

Moreover, to support robust evaluation on diverse acoustic conditions, we introduce \RaonEval, a structured benchmark constructed from 12 publicly available speech datasets, including CMU-ARCTIC~\citep{kominek04b_ssw}, TED-LIUM3~\citep{hernandez2018ted}, and EmoV-DB~\citep{adigwe2018emotional}. We group these datasets into four acoustic regimes, Clean, Noisy, Wild, and Expressive to cover controlled read speech, noisy environments, in-the-wild conversations, and expressive speech. 
Unlike existing benchmarks that rely on prompts drawn from a single read-speech dataset and evaluate only a limited number of generations, \RaonEval comprises 6K speech prompt-text pairs sampled from diverse datasets, enabling robust assessment across acoustic conditions.
On \RaonEval, \RaonModel-1B achieves the best average WER and SIM across acoustic regimes while matching the top-performing models in human preference scores, such as comparative mean opinion score (CMOS) and similarity mean opinion score (SMOS).

Our contributions are summarized as follows:

\begin{itemize}
  \item We introduce \textbf{\RaonPool}, a 615K-hour English speech corpus aggregated from 10 publicly available sources and web-sourced recordings, which is the largest openly available multi-source dataset curated for TTS training to our knowledge.

  \item We derive \textbf{\RaonCore}, a filtered subset of 510K hours and 194M speech segments from \RaonPool, and show that training on \RaonCore consistently improves zero-shot TTS performance over the unfiltered pool.

  \item We introduce \textbf{\RaonEval}, a robustness-oriented evaluation benchmark spanning four acoustic regimes (Clean, Noisy, Wild, Expressive) with 6K prompt-text
  pairs, enabling systematic analysis of zero-shot TTS across diverse real-world conditions.
  
  \item We release \textbf{\RaonModel}, a DiT-based TTS model family (0.3B, 1B parameters), which ranks first or second in WER and SIM among recent zero-shot TTS baselines on Seed-TTS-Eval and CV3-Eval, competitive with models trained on large-scale proprietary data.

\end{itemize}

\begin{table}[t]
\centering
\caption{Results on Seed-TTS-Eval~\citep{anastassiou2024seed} benchmark. {\RaonModel} outperforms the open-data, open-weight baselines while remaining comparable with strong open-weight, closed-data TTS baselines. We report the word error rate~(WER) and speaker similarity~(SIM). \textbf{Bold} indicates the best performance, and \underline{underline} indicates the second-best performance among the compared open-weight models.} 

\label{tab:seed-eval}
\begin{tabular}{lcccccc}
\toprule[1.2pt]
{Model} 
& {Param.} 
& {Training Data} 
& {Open-Weight} 
& {Open-Data} 
& {WER}$\downarrow$ 
& {SIM}$\uparrow$ \\
\midrule[0.4pt]

{Human} & - & - & - & - & 2.14 & 0.734 \\
\midrule[0.4pt]
\rowcolor{gray!15}
{Seed-TTS} & - & - & & & 2.25 & 0.762 \\
\rowcolor{gray!15}
{CosyVoice~3} & 1.5B & $\sim$ 1M & & & 2.21 & 0.720 \\

\midrule[0.4pt]
{Index-TTS 2} & 1.5B & 55K & \cmark &  & 2.18 & 0.709 \\
{Llasa} & 8B & 250K & \cmark &  & 3.63 & 0.581 \\
{VoxCPM} & 0.5B & 1.8M & \cmark &  & 1.98 & \underline{0.730} \\
{CosyVoice~2} & 0.5B & 170K & \cmark &  & 2.61 & 0.659 \\
{CosyVoice~3} & 0.5B & $\sim$ 1M & \cmark & & 2.50 & 0.698 \\
{Qwen3-TTS} & 1.7B & $\sim$ 5M & \cmark & & {\textbf{1.46}} & 0.715 \\
{Voxtral TTS} & 4B & - & \cmark & & 2.19 & 0.663 \\
\midrule

{MaskGCT} & 0.6B & 100K & \cmark & \cmark & 2.57 & 0.713 \\
F5-TTS & 0.3B & 100K & \cmark & \cmark & 2.04 & 0.671 \\
\midrule[0.4pt]
\RaonModel-0.3B & 0.3B & 510K & \cmark & \cmark & 1.95 & 0.687 \\
\RaonModel-1B & 1.0B & 510K & \cmark & \cmark & \underline{1.78} & \textbf{0.749} \\
\bottomrule[1.2pt]
\end{tabular}
\end{table}

\section{Background \& Related Work}
\label{sec:related_work}

\subsection{Text-to-speech models}

\noindent\textbf{Discrete approaches} generate speech by modeling sequences of discrete acoustic or codec tokens.
VALL-E~\citep{wang2023neuralcodeclanguagemodels} introduced the neural codec language modeling paradigm, framing zero-shot TTS as conditional token prediction over discrete codec representations.
Building on this, MaskGCT~\citep{wang2024maskgct} applies masked generative modeling for efficient zero-shot synthesis, VoxCPM~\citep{zhou2025voxcpm} proposes a tokenizer-free formulation improving context modeling and voice fidelity, and Seed-TTS~\citep{anastassiou2024seed} demonstrates that scaling such models enables robust generation across diverse speakers and conditions.
More recent systems extend this paradigm with large language model architectures: LLaSA~\citep{ye2025llasa} scales codec language modeling to multi-billion parameters, and Qwen3-TTS~\citep{hu2026qwen3} adopts an LLM-based architecture over discrete speech tokens for high-quality synthesis.

\noindent\textbf{Continuous approaches} generate speech directly in continuous acoustic or latent spaces, typically using diffusion or flow-based objectives. 
DiTTo-TTS~\citep{lee2024ditto} employs a DiT-based latent diffusion framework that generates speech in compressed latent representations.
F5-TTS~\citep{chen2025f5} adopts a DiT backbone with a flow-matching objective, enabling efficient speech synthesis with simplified text-speech alignment.

\noindent\textbf{Hybrid approaches} combine discrete semantic representations with continuous acoustic generation to improve controllability and speech quality. 
CosyVoice~2 and~3~\citep{du2024cosyvoice,du2025cosyvoice} integrate large language model-based semantic token generation with conditional flow matching for acoustic synthesis, enabling multilingual and instruction-following capabilities. 
IndexTTS~2~\citep{zhou2026indextts2} explores semantic indexing for efficient style manipulation and controllable speech generation. 
Voxtral-TTS~\citep{liu2026voxtral} adopts a unified LLM-based architecture that jointly models semantic and acoustic aspects of speech, with acoustic generation formulated using a flow-matching objective.

\subsection{Open datasets for foundation models}

\noindent\textbf{Text (LLMs).}
Open datasets have played a central role in enabling reproducible LLM training.
The Pile~\citep{gao2020pile} established an early publicly available resource for LLM pretraining, and subsequent efforts, including RefinedWeb~\citep{penedo2023refinedweb}, RedPajama~\citep{weber2024redpajama}, Dolma~\citep{soldaini2024dolma}, and FineWeb~\citep{penedo2024fineweb}, improved the scale and quality of open web-derived text corpora.
DataComp-LM~\citep{li2024datacomp} further systematically studied data curation strategies and scaling behavior for open LLM training.

\noindent\textbf{Image-text models.}
In vision and multimodal learning, LAION~\citep{schuhmann2022laion} provides billions of web-mined image-text pairs, and DataComp~\citep{gadre2023datacomp} introduces a controlled benchmark for studying dataset curation at scale.
These resources have supported OpenCLIP~\citep{cherti2023reproducible}, which reproduces CLIP~\citep{radford2021learning} using entirely open data.

\noindent\textbf{Speech (ASR and TTS).}
Similar efforts have emerged in speech recognition to reproduce Whisper~\citep{radford2022robustspeechrecognitionlargescale}, which is trained on proprietary data.
OWSM~\citep{peng2024owsm} investigates Whisper-style training using public datasets and open-source toolchains, and OLMoASR~\citep{ngo2025olmoasr} introduces a large-scale open dataset and curation pipeline for fully reproducible ASR training.
Despite these advances, reproducible large-scale training remains underexplored in TTS.
Compared to ASR, TTS requires more reliable utterance-level speech-text alignment and consistent speaker characteristics, making large-scale data construction substantially more challenging. Existing open TTS datasets fall short of the multi-million-hour scale used by many recent TTS systems, whose training data is not publicly available.

\section{Methods}
\label{sec:method}\begin{figure}[t]
    \centering    \includegraphics[width=\linewidth,trim={3cm 4cm 2cm 5cm},height=0.6\textheight,keepaspectratio]{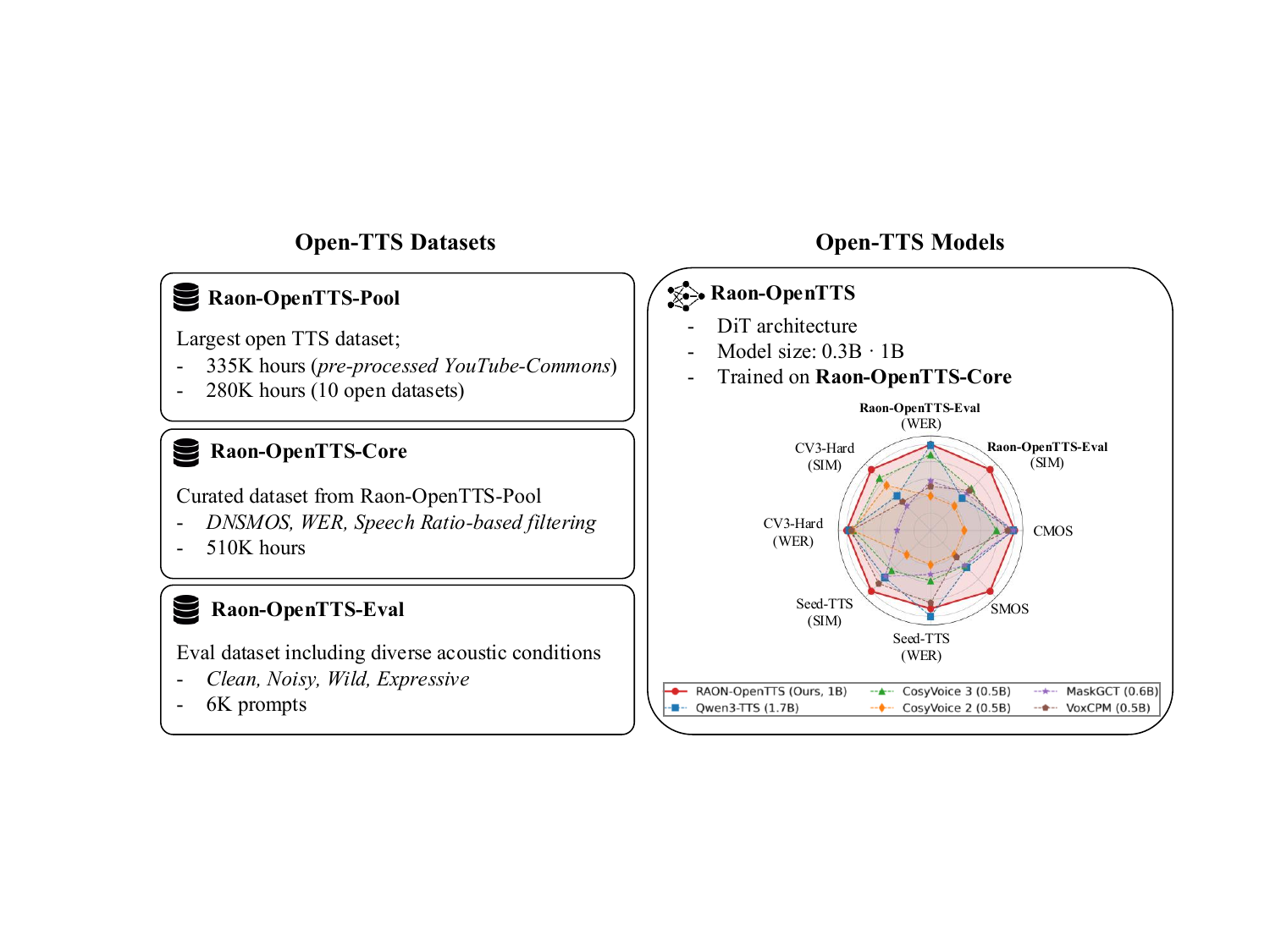}
    \caption{\textbf{Overview of \RaonModel.} \RaonPool contains 615K hours of English speech, including 335K hours from pre-processed YouTube-Commons and 280K hours from 10 publicly available datasets. \RaonCore is a 510K-hour filtered subset constructed using Deep Noise Suppression Mean Opinion Score (DNSMOS), WER, and Speech Ratio-based filtering. \RaonEval is a robustness-oriented benchmark with 6K prompts across Clean, Noisy, Wild, and Expressive conditions. \RaonModel is a DiT-based TTS model family with 0.3B and 1B variants trained on \RaonCore. The radar chart summarizes objective performance, measured by WER and SIM on Seed-TTS, CV3-Hard, and RAON-OpenTTS-Eval, together with subjective evaluation results based on comparative mean opinion score (CMOS) and similarity mean opinion score (SMOS).}
    \label{fig:raon_overview}
\end{figure}

In this section, we introduce \RaonPool, a large-scale open TTS data pool collected from diverse sources, \RaonCore, a curated, high-quality subset of \RaonPool, and \RaonModel, our series of open-weight TTS models trained on the proposed datasets.
Figure~\ref{fig:raon_overview} summarizes the overall data collection, curation, and training pipeline of the \RaonModel family.

\subsection{\RaonPool}\label{sec:pool}

We construct \RaonPool, a large-scale open English speech corpus by aggregating diverse public datasets.
Constructing such a dataset is non-trivial, as TTS requires not only transcribed speech but also reliable utterance-level alignment and speaker consistency.
Accordingly, we combine both TTS-oriented and ASR-style datasets to achieve both scale and diversity.

\noindent\textbf{Data Sources.} We restrict our data sources to publicly available English speech datasets with more than 500 hours of audio, as smaller corpora contribute only marginally to the overall data volume. In addition, including many small datasets increases dataset fragmentation and management complexity. We further limit speech segments to those shorter than 30 seconds to reduce alignment errors, multi-speaker content, and non-speech artifacts. Since Seed-TTS-Eval includes samples derived from Common Voice, we exclude all Common Voice data when training the TTS models used for evaluation to ensure fair comparison.
Based on these criteria, \RaonPool is constructed by combining and processing 11 publicly available English speech-text corpora, including LibriTTS-R~\citep{koizumi2023libritts}, SPGISpeech~\citep{o2021spgispeech}, SPGISpeech 2.0~\citep{grossman2025spgispeech}, HiFiTTS2~\citep{langman2025hifitts}, LibriHeavy~\citep{kang2024libriheavy}, GigaSpeech~\citep{chen2021gigaspeech}, VoxPopuli~\citep{wang2021voxpopuli}, People’s Speech~\citep{galvez2021people}, Emilia~\citep{he2024emilia}, Emilia-YODAS~\citep{he2024emilia}, and YouTube-Commons~\footnote{https://huggingface.co/datasets/PleIAs/YouTube-Commons}. 

Four of the constituent datasets are originally multilingual corpora; we extract only the English subset from each.
For Emilia and Emilia-YODAS, we use the official English split. For VoxPopuli, we select English sessions from the European Parliament recordings. For YouTube-Commons, we first select videos labeled as English in the original metadata, then verify language identity using Whisper-based language detection during preprocessing.
The collection spans both TTS-oriented and ASR datasets; ASR corpora are included when they provide utterance-level aligned speech-text pairs, as they broaden speaker, domain, and acoustic diversity.
For VoxPopuli, where complete transcriptions are unavailable or partially missing, we construct speech-text pairs by referencing available transcripts from MoSEL~\citep{gaido2024mosel}.

As shown in Table~\ref{tab:data-statistics}, these datasets span a wide range of domains, recording conditions, and annotation qualities, and are 
released under the following licenses: CC BY~4.0, CC-0, Apache~2.0, Public Domain, Kensho User Agreement (non-commercial research only) and CC BY-NC~4.0.
A substantial portion of \RaonPool is derived from YouTube-Commons, contributing 335K hours of speech data. 
Since the original release provides only YouTube URLs with noisy or unreliable transcriptions, we reconstruct this portion by collecting the audio and processing it into a high-quality speech-text dataset, which we refer to as \RaonYouTube and use as part of the training data.

\noindent\textbf{Preprocessing Pipeline for YouTube-Commons.}
Unlike the other datasets in \RaonPool, which are mostly released as utterance-level speech-text pairs, the \RaonYouTube dataset consists of long-form recordings ranging from approximately 15 minutes to several hours. A substantial portion of these recordings includes non-speech content, background audio, or low-quality speech that is unsuitable for TTS training, requiring preprocessing to extract reliable speech-text pairs. For the YouTube-Commons portion, we therefore apply a precise data preprocessing pipeline inspired by prior work~\citep{he2024emilia}. The pipeline consists of the following stages:

\begin{enumerate}[label=(\alph*)]
    \item Audio standardization and loudness normalization.
    All audio is resampled to 16~kHz mono and normalized to a fixed loudness level to reduce variability across sources.

    \item Source separation (UVR-MDX\footnote{https://github.com/Anjok07/ultimatevocalremovergui}).
    We apply UVR-MDX to suppress background music and non-vocal components, improving downstream diarization and ASR robustness.

    \item Speaker diarization (PyAnnote~3.1\footnote{https://huggingface.co/pyannote/speaker-diarization-3.1}).
    Speaker boundaries are estimated to ensure that each segment contains speech from a single dominant speaker.

    \item Voice activity detection (Silero VAD\footnote{https://github.com/snakers4/silero-vad}).
    Continuous speech regions are segmented into clips ranging from 3 to 30 seconds to balance context length and model stability.

    \item Automatic transcription (Whisper-large-v3\footnote{https://huggingface.co/openai/whisper-large-v3}).
    Each segment is transcribed using Whisper-large-v3 to obtain aligned speech-text pairs.
    
\end{enumerate}

We remove samples for which the transcription length is
disproportionately short relative to the corresponding audio duration, as these typically
indicate severe misalignment or transcription failures.
Finally, all audio from the collected datasets is standardized to a unified representation.
For storage efficiency, audio is stored in 64~kbps Opus format and is decompressed,
resampled to 16~kHz, and normalized for training.
\begin{table*}[t]
  \centering
  \small
  \caption{
    Data composition and quality statistics of {\RaonPool}. For each dataset, we report total data size, average duration and number of speech segments, license, and their quality scores, including Deep Noise Suppression Mean Opinion Score (DNSMOS; shown as DNS in the table), WER, and Speech Ratio (SR). We exclude speech segments longer than 30 seconds.
  }
  \label{tab:data-statistics}
  \setlength{\tabcolsep}{5pt}
  \begin{tabular}{l|cccl|ccc}
    \toprule
    {Dataset} &
    {Size(h)} &
    {Avg. Dur.(s)} &
    {Sgmts.(M)} &
    {License} &
    {DNS$\uparrow$} &
    {WER$\downarrow$} &
    {SR$\uparrow$} \\
    \midrule
    Raon-YouTube-Commons$^\dagger$ & 335K &  8.5 & 141.7 & CC BY 4.0     & 2.74 & 0.30 & 0.90 \\
    Emilia-YODAS$^\dagger$    & 92K  &  9.2 &  36.0 & CC BY-NC 4.0  & 2.82 & 0.19 & 0.90 \\
    Emilia$^\dagger$          & 47K  &  9.3 &  18.1 & CC BY 4.0     & 3.02 & 0.18 & 0.89 \\
    LibriHeavy      & 42K  & 14.2 &  10.8 & Public Domain & 3.22 & 0.11 & 0.83 \\
    HiFiTTS2        & 37K  & 10.1 &  13.1 & CC BY 4.0     & 3.20 & 0.11 & 0.84 \\
    People's Speech & 28K  & 14.2 &   7.0 & CC BY 4.0     & 2.63 & 0.25 & 0.86 \\
    VoxPopuli$^\dagger$       & 17K  & 27.8 &   2.2 & CC-0          & 2.82 & 0.36 & 0.83 \\
    GigaSpeech      & 10K  & 4.3  &   8.3 & Apache 2.0    & 2.73 & 0.16 & 0.90 \\
    SPGISpeech      & 5K  & 9.2  & 2.0 & Kensho UA$^\ddagger$ & 2.90 & 0.03 & 0.86 \\
    SPGISpeech 2.0     & 889  & 14.4  & 0.2 & Kensho UA & 2.72 & 0.08 & 0.90 \\
    LibriTTS-R      & 552  & 5.6  & 0.4 & CC BY 4.0 & 2.96 & 0.06 & 0.91 \\
    \midrule
    {Total / Avg.} & 615K & 9.2 & 239.7 & -- & 2.83 & 0.24 & 0.89 \\
    \bottomrule
    \multicolumn{8}{l}{\footnotesize $^\dagger$Originally multilingual; we use only the English subset.}\\
    \multicolumn{8}{l}{\footnotesize $^\ddagger$Kensho User Agreement (non-commercial research only).}                                                                                  
  \end{tabular}
\end{table*}

\label{sec:data-statistics}
\noindent\textbf{Data statistics.}
\RaonPool consists of 615K hours of speech audio and 240M speech segments.
Table~\ref{tab:data-statistics} summarizes both the scale and quality characteristics of each dataset in \RaonPool.
The constituent datasets exhibit a broad temporal coverage in average segment length. LibriTTS-R and GigaSpeech have relatively short average durations of 5.6 and 4.3 seconds, whereas VoxPopuli contains much longer segments with average durations of 27.8 seconds. We also report DNSMOS~\citep{reddy2021dnsmos}, WER, and Speech Ratio for each dataset.
WER is measured by transcribing each speech segment using a Whisper-small ASR model and computing the mismatch with the existing text annotations, while Speech Ratio is estimated using Silero VAD to quantify the proportion of frames containing active speech.
Audiobook-derived read-speech datasets such as HiFiTTS2 and LibriHeavy achieve relatively high DNSMOS scores of 3.20 and 3.22, respectively, along with a low WER of 0.11, indicating cleaner acoustic conditions.
In contrast, ASR-oriented datasets such as People’s Speech, GigaSpeech, and VoxPopuli, exhibit lower DNSMOS scores in the range of 2.63-2.82 and higher WER values between 0.16 and 0.36.
The \RaonYouTube subset shows a relatively high WER of 0.30, which reflects its fully in-the-wild nature, encompassing diverse recording environments, background noise, and spontaneous speech patterns.
This stands in contrast to datasets with highly reliable transcriptions, such as LibriTTS-R and SPGISpeech, which achieve much lower WER values (0.06 and 0.03, respectively).
These variations in acoustic quality and transcription reliability highlight the heterogeneous nature of \RaonPool, motivating systematic filtering to obtain a more suitable TTS training set.

\begin{figure}[t]
    \centering    \includegraphics[width=\linewidth,height=0.6\textheight,keepaspectratio]{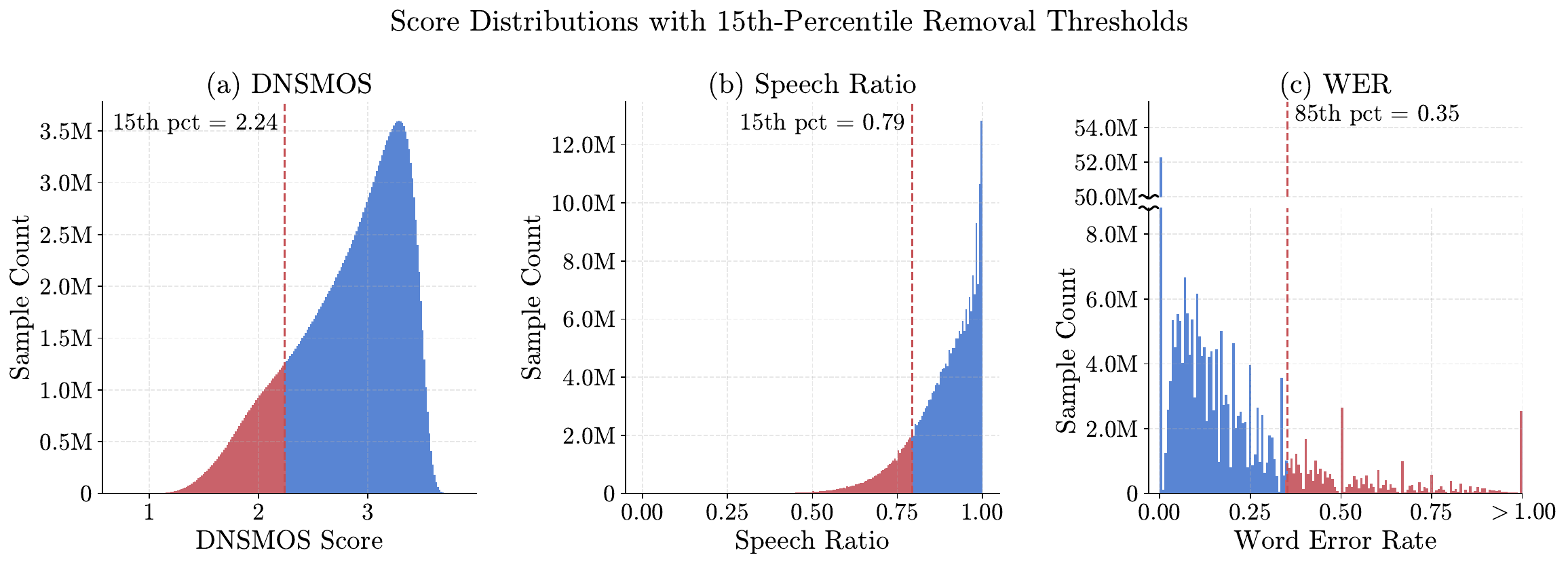}
    \caption{Distribution of data quality scores in \RaonPool and the percentile-based filtering thresholds used to construct \RaonCore. Histograms show the distributions of DNSMOS, Speech Ratio, and WER. The dashed vertical lines indicate the cutoff corresponding to removal of the lowest-quality 15\% tail for each metric, yielding thresholds of 2.24 for DNSMOS, 0.79 for Speech Ratio, and 0.35 for WER. These thresholds align with visually distinct low-quality tails while preserving the majority of the training data.}
    \label{fig:threshold_selection}
\end{figure}

\begin{table*}[t]
\centering
\caption{
Data filtering ablation results on Seed-TTS-Eval, CV3-Hard-EN, and \RaonEval (average over Clean/Noisy/Wild/Expressive subsets).
All models in this table use ablation-specific 0.3B variants trained for fewer steps with a different learning-rate schedule from the main \RaonModel-0.3B.
The final column reports the overall average rank across all evaluation metrics, where lower is better; DNS denotes DNSMOS, and Rank denotes average rank.
\textbf{Bold} indicates the best performance, and \underline{underline} indicates the second-best performance.
WER\,$>$\,100\% samples (Whisper hallucinations) are excluded from WER computation.
}
\label{tab:filtering_ablation}
\small
\setlength{\tabcolsep}{4pt}
\begin{tabular}{l ccc c ccc ccc c}
\toprule
 & \multicolumn{3}{c}{Seed-TTS-Eval}
 & \multicolumn{1}{c}{CV3-EN}
 & \multicolumn{3}{c}{CV3-Hard-EN}
 & \multicolumn{3}{c}{\RaonEval}
 & \multicolumn{1}{c}{Overall} \\
\cmidrule(lr){2-4} \cmidrule(lr){5-5} \cmidrule(lr){6-8} \cmidrule(lr){9-11} \cmidrule(lr){12-12}
Filtering
 & WER$\downarrow$ & SIM$\uparrow$ & DNS$\uparrow$
 & WER$\downarrow$
 & WER$\downarrow$ & SIM$\uparrow$ & DNS$\uparrow$
 & WER$\downarrow$ & SIM$\uparrow$ & DNS$\uparrow$
 & Rank$\downarrow$ \\
\midrule
Combined (15\%) & 2.00 & \textbf{0.672} & 3.12 & \underline{4.25} & 8.14 & \textbf{0.642} & 3.11 & 4.46 & \textbf{0.611} & \textbf{3.05} & \textbf{3.40} \\
DNSMOS (15\%) & \textbf{1.97} & 0.669 & \textbf{3.15} & \textbf{4.18} & \underline{7.33} & 0.617 & \textbf{3.15} & 4.67 & 0.602 & 3.00 & \underline{3.60} \\
WER (15\%) & 1.99 & 0.668 & \underline{3.15} & 4.47 & 8.20 & 0.620 & 3.12 & \textbf{3.66} & 0.594 & 3.03 & 4.50 \\
DNSMOS (50\%) & \underline{1.98} & 0.671 & 3.14 & 4.90 & \textbf{7.20} & 0.622 & 3.13 & 5.34 & 0.588 & 3.02 & 4.70 \\
No filtering & 2.19 & 0.661 & 3.14 & 4.73 & 8.53 & 0.628 & 3.12 & 4.30 & 0.603 & 3.03 & 4.90 \\
Combined (50\%) & 2.32 & \underline{0.672} & 3.14 & 5.01 & 7.56 & \underline{0.630} & \underline{3.15} & 4.83 & 0.601 & 3.01 & 4.90 \\
VAD (15\%) & 2.22 & 0.665 & 3.13 & 4.81 & 7.86 & 0.621 & 3.11 & \underline{4.24} & \underline{0.604} & \underline{3.04} & 5.20 \\
VAD (50\%) & 2.10 & 0.666 & 3.10 & 4.27 & 7.69 & 0.620 & 3.11 & 4.58 & 0.597 & 2.99 & 6.20 \\
WER (50\%) & 2.20 & 0.655 & 3.14 & 5.17 & 9.88 & 0.607 & 3.11 & 4.96 & 0.590 & 3.03 & 7.60 \\
\bottomrule
\end{tabular}
\end{table*}

\subsection{\RaonCore}\label{sec:raoncore}

In this section, we describe the construction of \RaonCore, a curated subset of \RaonPool designed to improve data quality for TTS training. 
While \RaonPool provides large-scale coverage, it contains substantial variation in acoustic quality and transcription reliability, which can negatively affect model performance. 
To address this, we filter out low-quality samples using three model-based criteria: WER, Deep Noise Suppression Mean Opinion Score (DNSMOS), and Speech Ratio, resulting in a more reliable training set.
\begin{enumerate}[label=(\alph*)]
    \item WER-based filtering.
    We transcribe each audio segment using a Whisper-small ASR model and compute the WER against the existing text annotation.
    Segments with excessively high WER are discarded, as they are likely to contain severe transcription mismatches or highly degraded speech.

    \item DNSMOS-based filtering.
    We estimate perceptual speech quality using DNSMOS and remove segments falling below a minimum quality threshold, targeting recordings with strong background noise or distortion.

    \item VAD-based filtering.
    We estimate speech segment proportion of each segment, denoted Speech Ratio, using Silero VAD and discard segments with an unusually low Speech Ratio, which typically correspond to recordings dominated by silence, music, or non-speech audio.

    \item Combined filtering.
    We jointly consider all three quality signals by computing an absolute rank for each segment along each criterion (DNSMOS, WER, and SR) and averaging the ranks into a single combined score. We then discard segments falling below a 15th-percentile threshold. This approach prevents any single noisy signal from dominating the filtering decision and yields more stable performance across diverse evaluation settings. (see Table~\ref{tab:filtering_ablation}).  

\end{enumerate}

\noindent\textbf{Threshold selection.}
We adopt a 15th-percentile removal threshold for each quality metric, where the percentile is computed over speech segments rather than total audio hours. As shown in Figure~\ref{fig:threshold_selection}, this percentile lies near the low-quality tail of each empirical distribution rather than the main body. For DNSMOS, this corresponds to values below 2.24, indicating acoustically degraded recordings. For SR, most samples are concentrated near 0.9 to 1.0, and values below 0.79 correspond to segments containing more than 21\% non-speech content. For WER, the distribution is sharply concentrated near zero, while values above 0.35 are associated with severely mismatched or unreliable speech-text pairs.
This threshold represents a conservative balance between removing clearly low-quality samples and retaining sufficient data scale and diversity. We additionally consider a more aggressive 50\% removal threshold to assess the effect of over-filtering on model performance.
  \begin{table}[t]
  \centering
  \caption{Per-dataset retention after combined 15th-percentile filtering.
  Retention denotes the fraction of segments preserved in \RaonCore.
  Studio-quality corpora retain over 93\%, while noisier ASR-oriented
  datasets are more aggressively filtered.
  Notably, \RaonYouTube retains 82.9\% of its segments despite its
  in-the-wild origin, indicating that the majority of its data meets
  the quality threshold for TTS training.}
  \label{tab:core-retention}
  \begin{tabular}{l r r r}
  \toprule
  Dataset & Core (M) & Retention (\%) \\
  \midrule
  LibriTTS-R           &   0.3 & 97.7 \\
  HiFiTTS2             &  11.9 & 94.5 \\
  LibriHeavy           &  10.2 & 94.4 \\
  Emilia               &  16.8 & 93.2 \\
  Emilia-YODAS         &  31.6 & 87.8 \\
  People's Speech (Clean) & 1.3 & 83.9 \\
  Raon-YouTube-Commons & 117.5 & 82.9 \\
  VoxPopuli            &   1.6 & 71.8 \\
  People's Speech (Dirty) & 2.6 & 48.2 \\
  \midrule
  Total                & 193.9 & 84.7 \\
  \bottomrule
  \end{tabular}
  \end{table}

\paragraph{Filtering ablation.}
To investigate the effectiveness of the filtering pipeline, we conduct an ablation study using \RaonModel-0.3B (see Section~\ref{sec:model} for details). 
All models are trained with a fixed number of update steps corresponding to one epoch of the full dataset (i.e., the unfiltered \RaonPool), ensuring that performance differences arise from data quality rather than training duration.
Table~\ref{tab:filtering_ablation} summarizes the results across multiple benchmarks. Among all strategies, combined filtering with a 15th-percentile removal threshold achieves the best average rank, indicating the most consistent overall performance. 
In general, moderate filtering (15\%) outperforms more aggressive filtering (50\%), suggesting that removing only the lowest-quality tail improves data quality while preserving sufficient diversity. 
Table~\ref{tab:core-retention} shows the per-dataset effect of the combined filter. Studio-quality read-speech datasets such as LibriTTS-R (97.7\%), HiFiTTS2 (94.5\%), and LibriHeavy (94.4\%) are nearly fully preserved, consistent with their high DNSMOS and low WER (Table~\ref{tab:data-statistics}). In contrast, People's Speech (Dirty) retains only 48.2\% due to its lower DNSMOS (2.63) and higher WER (0.25). Notably, \RaonYouTube retains 82.9\% of its segments despite its in-the-wild recording conditions, suggesting that the preprocessing pipeline (Section~\ref{sec:pool}) effectively produces segments of sufficient quality for TTS training.

\subsection{\RaonModel}
\label{sec:model}
\paragraph{Model architecture.}
Our model is based on the DiT architecture introduced in F5-TTS for zero-shot TTS synthesis. We intentionally adopt the original architecture without alteration in order to isolate the effect of data scale and coverage. Based on this architecture, we train three model variants of different scales: \textbf{\RaonModel-0.3B} and \textbf{\RaonModel-1B}, both trained on \textbf{\RaonCore}.  All models share the same architectural design and training objective, while differing in model capacity through scaling the number of transformer layers, attention heads, attention embedding dimension, and feed-forward dimension, as summarized in Appendix~\ref{sec:model_config}.

\paragraph{Training configurations.}
  All models are trained using distributed training on NVIDIA B200 GPUs with gradient norm clipping at a maximum norm of 1.0.                   
  \RaonModel-0.3B is trained for 225K update steps with a per-GPU batch size of 35K audio frames, a peak learning rate of $7.5\times10^{-5}$    
  linearly warmed up for 50K steps and linearly decayed for the remainder.   
  \RaonModel-1B is trained for 550K update steps with a per-GPU batch size of 14K audio frames, a peak learning rate of $1\times10^{-4}$ linearly warmed up for 50K steps and linearly decayed for the remainder.   
  \RaonModel-0.3B and \RaonModel-1B require approximately 1K and 9K GPU-hours for training, respectively.
  Audio is represented using 80-channel log mel-spectrogram features extracted at a sampling rate of 16~kHz with a hop size of 256. Text is
  represented at the character level, with a vocabulary size of 5,512. For waveform synthesis, we use the HiFi-GAN\footnote{https://huggingface.co/speechbrain/tts-hifigan-libritts-16kHz} vocoder pretrained on LibriTTS at 16~kHz.
  For inference, both \RaonModel and F5-TTS use 32 NFE steps for ODE sampling.
  The 0.3B models used in the filtering and data ablation studies are separate ablation variants trained for fewer steps with a different learning-rate schedule.

\section{Experimental setup}
\label{sec:experiment}

\subsection{Evaluation on conventional benchmarks}
We employ two representative benchmarks for zero-shot TTS evaluation: Seed-TTS-Eval and
CV3-Eval. For Seed-TTS-Eval, we use the English subset (Seed-TTS-EN) only.
This benchmark reports intelligibility measured by WER and SIM.
WER is computed by transcribing synthesized speech using Whisper-large-v3~\citep{radford2022robustspeechrecognitionlargescale}, while SIM is measured
using speaker embeddings extracted from a WavLM-large~\citep{chen2022wavlm} fine-tuned for speaker verification,
following the official evaluation protocol.
For CV3-Eval, we evaluate on both the English (EN) and hard English (Hard-EN)
subsets.
On the EN subset, intelligibility is evaluated using WER.
The more challenging Hard-EN subset contains linguistically complex English sentences and
is evaluated using three complementary metrics: WER for intelligibility, SIM for speaker similarity,
and DNSMOS for perceived speech quality.
For speaker similarity on Hard-EN, we follow the official protocol and use an ERes2Net~\citep{chen2023enhanced}-based
speaker verification model.

\subsection{\RaonEval}\label{sec:raon-eval}
Existing zero-shot TTS benchmarks typically evaluate models using around 1,000 prompts drawn from a single read-speech dataset to assess zero-shot synthesis ability in a target language.
As a result, the evaluation is tied to the specific acoustic characteristics of the dataset, providing an incomplete view of model robustness, particularly under realistic and challenging recording scenarios. 
Moreover, prior benchmarks rarely analyze robustness in a structured manner across different acoustic regimes.
To address these limitations, we introduce \RaonEval, a new benchmark designed to enable a more systematic robustness analysis of zero-shot TTS models. 
We categorize prompt speech into four acoustic categories according to recording conditions, speaking style, and acoustic variability: Clean, Noisy, Wild, and Expressive.
Across these categories, \RaonEval comprises a total of 12 evaluation datasets covering controlled read speech, noisy and reverberant environments, spontaneous conversational speech, and expressive speech as follows:

\paragraph{Clean.}
The Clean category consists of high-quality, studio-like read speech recorded under
controlled acoustic conditions.
These datasets serve as a reference setting for evaluating pronunciation accuracy and
naturalness under minimal acoustic distortion.
We include LibriSpeech-clean \citep{panayotov2015librispeech}, ST American English \footnote{https://www.openslr.org/45/}, CMU-ARCTIC \citep{kominek04b_ssw}, L2-ARCTIC \citep{zhao18b_interspeech}, and VCTK \citep{yamagishi2019cstr}, totaling five datasets.

\paragraph{Noisy.}
The Noisy category includes read or prompted speech recorded in the presence of
noticeable background noise or reverberation.
These datasets evaluate robustness to environmental interference, channel variability,
and mild acoustic corruption.
We use LibriSpeech-other and TED-LIUM~3 \citep{hernandez2018ted} as representative noisy-speech benchmarks.

\paragraph{Wild.}
The Wild category contains unscripted or conversational speech captured in
uncontrolled real-world settings and subsequently transcribed.
These datasets test robustness to spontaneous speech phenomena and
real-world conversational recording conditions that are rarely covered
by clean-speech benchmarks.
This category includes AMI-IHM and AMI-SDM~\citep{kraaij2005ami}. 
For AMI-SDM, we observe a substantial number of samples with noisy or misaligned transcriptions. To ensure reliable evaluation, we retain only segments with zero word error rate (WER), as estimated by a Whisper ASR model, filtering out samples with severe transcription mismatches.

\paragraph{Expressive.}
The Expressive category focuses on speech with explicit affective content,
covering a wide range of emotional expressions and prosodic styles.
These datasets assess a model’s ability to generalize beyond neutral speaking styles
and controlled prosody.
We include Expresso \citep{nguyen2023expresso}, CREMA-D \citep{cao2014crema}, and EmoV-DB \citep{adigwe2018emotional}.

For each evaluation dataset, we sample 500 utterances as speech prompts, selected via stratified sampling when speaker metadata is available to ensure balanced coverage across speakers and characteristics such as emotion, dialect, and speaking style.
Each prompt is paired with a target text drawn from a disjoint sentence in the same dataset.
Across the 12 datasets, this protocol yields 6,000 prompt-text pairs, including 2,500 from Clean, 1,000 from Noisy, 1,000 from Wild, and 1,500 from Expressive.
For evaluation, we report normalized WER, computed after applying the official Whisper text normalizer to both reference and hypothesis transcripts to avoid penalizing surface-form variations such as numeric expressions or hyphenated compounds.
We also report SIM, measured as cosine similarity between WavLM speaker embeddings of the prompt and generated audio.

\subsection{Baselines}
We compare \RaonModel against 9 recent zero-shot TTS models.
For all open-weight baseline models, we use their officially released checkpoints and follow the default or recommended inference settings provided by the authors, without additional tuning.
For CosyVoice~3, we use the officially released 0.5B checkpoint, as the 1.5B variant is not available as open weights.
For closed-weight models, we report the evaluation results as published by the original authors, since direct inference with these models is not publicly accessible.
Our evaluation includes Seed-TTS~\citep{anastassiou2024seed},  CosyVoice~2~\citep{du2024cosyvoice}, CosyVoice~3~\citep{du2025cosyvoice}, Index-TTS~2~\citep{zhou2026indextts2}, 
Llasa~\citep{ye2025llasa},
VoxCPM~\citep{zhou2025voxcpm}, MaskGCT~\citep{wang2024maskgct}, F5-TTS~\citep{chen2025f5}, and Qwen3-TTS~\citep{hu2026qwen3}.

\section{Results}
\label{sec:results}\begin{table}[t]
\centering
\caption{Results on the CV3-EN and CV3-Hard-EN zero-shot TTS benchmarks.
\textbf{Bold} indicates the best performance, and \underline{underline} indicates the second-best performance. F5-TTS fails to generate most CV3-Hard-EN samples and 15 out of 500 CV3-EN samples due to its inability to handle long text inputs; these are excluded from its results.}
\label{tab:cv3-eval}
\begin{tabular}{l c c c c}
\toprule
{Model} 
& \multicolumn{1}{c}{{CV3-EN}} 
& \multicolumn{3}{c}{{CV3-Hard-EN}} \\
\cmidrule(lr){2-2}
\cmidrule(lr){3-5}
& {WER}$\downarrow$
& {WER}$\downarrow$
& {SIM}$\uparrow$
& {DNSMOS}$\uparrow$ \\
\midrule

F5-TTS              & 8.54  & -   & -  & -  \\
MaskGCT          & 7.73  & 41.09 & 0.624   & 3.48   \\
CosyVoice~2          & 6.27 & 10.28 & 0.710 & \underline{3.95} \\
CosyVoice 3    & 4.96  & 10.77  & \underline{0.740} & \textbf{3.98} \\
VoxCPM    & 5.24 & \underline{6.44} & 0.670 & 3.78 \\
Qwen3-TTS    & \underline{4.52} & 7.89 & 0.666 & 3.87 \\
\midrule
\RaonModel-0.3B   & 4.62 & 7.31 & 0.730 & 3.77 \\
\RaonModel-1B     & \textbf{3.92} & \textbf{6.15} & \textbf{0.775} & 3.85 \\
\bottomrule
\end{tabular}
\end{table}

\subsection{Seed-TTS-Eval}
Table~\ref{tab:seed-eval} reports the results on the Seed-TTS-Eval~\citep{anastassiou2024seed}.
\RaonModel-1B achieves the second-best WER among all evaluated models, including closed-weight systems,
with a WER of 1.78\%, surpassed by Qwen3-TTS, which is trained on multi-million-hour proprietary corpora.
In terms of speaker similarity, \RaonModel-1B attains a score of 0.749, ranking first among all open-weight models and remaining competitive with larger or closed systems. 
The 0.3B model already outperforms F5-TTS, its same-size, same-architecture baseline, on both WER (1.95 vs.\ 2.04) and SIM (0.687 vs.\ 0.671), indicating that the gains of \RaonModel are not solely due to model scale.

\begin{table*}[t]
\centering
\caption{Comparison with recent state-of-the-art zero-shot TTS models under the proposed evaluation protocol.
Overall scores are computed over all evaluation samples across the four acoustic categories.
WER ($\downarrow$) and SIM~($\uparrow$) are reported.
\textbf{Bold} indicates the best performance, and \underline{underline} indicates the second-best performance.}
\vspace{5pt}
\label{tab:olmo-eval}
\setlength{\tabcolsep}{4pt}
\resizebox{\textwidth}{!}{
\begin{tabular}{l cc cc cc cc cc}
\toprule
\multirow{2}{*}{{Model}} &
\multicolumn{2}{c}{{Clean}} &
\multicolumn{2}{c}{{Noisy}} &
\multicolumn{2}{c}{{Wild}} &
\multicolumn{2}{c}{{Expressive}} &
\multicolumn{2}{c}{{Overall}} \\
\cmidrule(lr){2-3} \cmidrule(lr){4-5} \cmidrule(lr){6-7} \cmidrule(lr){8-9} \cmidrule(lr){10-11}
 &
{WER}$\downarrow$ & {SIM}$\uparrow$ &
{WER}$\downarrow$ & {SIM}$\uparrow$ &
{WER}$\downarrow$ & {SIM}$\uparrow$ &
{WER}$\downarrow$ & {SIM}$\uparrow$ &
{WER}$\downarrow$ & {SIM}$\uparrow$ \\
\midrule
F5-TTS   &
2.17 & 0.613 &
3.82 & 0.640 &
136.03 & 0.324 &
3.46 & 0.503 &
25.08 & 0.542 \\
MaskGCT        &
3.39 & 0.672 &
5.56 & 0.727 &
28.00 & 0.581 &
6.44 & 0.546 &
8.61 & 0.635 \\
CosyVoice~2     &
2.59 & 0.642 &
4.39 & 0.675 &
49.73 & 0.535 &
3.66 & 0.536 &
11.02 & 0.603 \\
CosyVoice~3     &
2.53 & 0.678 &
3.69 & 0.720 &
8.31 & \underline{0.618} &
5.49 & 0.567 &
4.43 & \underline{0.647} \\
VoxCPM         &
2.24 & \underline{0.686} &
\textbf{3.42} & \underline{0.738} &
43.83 & 0.553 &
\underline{2.66} & 0.565 &
9.48 & 0.642 \\
Qwen3-TTS         &
3.38 & 0.684 &
4.60 & 0.726 &
79.14 & 0.528 &
5.81 & 0.527 &
17.59 & 0.626 \\
\midrule
\RaonModel-0.3B & \underline{1.57} & 0.645 & 4.03 & 0.700 & \underline{5.83} & 0.571 & \textbf{2.53} & \underline{0.570} & \underline{2.93} & 0.623 \\
\RaonModel-1B & \textbf{1.44} & \textbf{0.718} & \underline{3.51} & \textbf{0.769} & \textbf{5.61} & \textbf{0.656} & 2.77 & \textbf{0.633} & \textbf{2.81} & \textbf{0.695} \\
\bottomrule
\end{tabular}
}
\end{table*}

\subsection{CV3-Eval}
Table~\ref{tab:cv3-eval} summarizes the results on CV3-EN~\citep{du2025cosyvoice} and CV3-Hard-EN.
\RaonModel-1B achieves strong performance on both evaluation settings, recording the lowest WER of
3.92 on CV3-EN and the lowest WER of 6.15 on the more challenging CV3-Hard-EN split, which consists of linguistically complex target texts.
In addition, \RaonModel-1B attains the highest SIM of 0.775 on CV3-Hard-EN,
indicating superior preservation of speaker similarity.
In terms of perceptual quality, \RaonModel-1B achieves a DNSMOS score of 3.85, which is comparable to
the best-performing system (CosyVoice~3 at 3.98).
Notably, unlike VoxCPM, which exhibits competitive performance on Seed-TTS-Eval but a
noticeable drop in SIM under the harder evaluation setting, \RaonModel-1B maintains strong SIM across both benchmarks.
Taken together, these results show that \RaonModel-1B delivers the strongest overall performance on CV3-Eval in intelligibility and speaker similarity, while remaining competitive in perceptual quality.
At the 0.3B scale, \RaonModel-0.3B also substantially outperforms F5-TTS, reducing CV3-EN WER from 8.54 to 4.62 and remaining evaluable on CV3-Hard-EN, where F5-TTS fails on most samples.

\subsection{\RaonEval}
Table~\ref{tab:olmo-eval} reports results on \RaonEval.
\RaonModel-1B demonstrates strong overall intelligibility on \RaonEval, with especially large gains in the Clean and Wild conditions. The large margin observed in the Wild split highlights the robustness of \RaonModel under highly unconstrained, real-world acoustic environments. \RaonModel-1B also achieves competitive WER values of 3.51 in Noisy and 2.77 in Expressive settings, indicating stable performance under both environmental noise and expressive speaking styles. Across all conditions, \RaonModel-1B maintains consistently strong speaker similarity, with an overall SIM score of 0.695 and no pronounced degradation even in the most challenging acoustic scenarios. These results suggest that \RaonModel-1B generalizes well across a wide range of realistic acoustic conditions while preserving speaker similarity in zero-shot synthesis.
Raon-OpenTTS-0.3B achieves an overall WER of 2.93, dramatically improving over F5-TTS (25.08), with especially large gains in the Wild condition (5.83 vs.\ 136.03). Despite its smaller size, it also achieves lower overall WER than several larger open-weight baselines, although its speaker similarity remains below the strongest larger models.
Interestingly, several autoregressive baseline models such as CosyVoice~2, Qwen3-TTS, and VoxCPM exhibit substantially higher WER in the Wild split. Such behavior is less apparent on conventional clean-speech benchmarks, highlighting the importance of evaluating zero-shot TTS models under diverse acoustic regimes.

\begin{table}[t]
\centering
\caption{Subjective evaluation results using comparative mean opinion score (CMOS), which measures perceived naturalness relative to a reference.
Scores are reported relative to \RaonModel-1B. \textbf{Bold} indicates the best performance, and \underline{underline} indicates the second-best performance.}
\label{tab:cmos}
\begin{tabular}{l c c c c c}
\toprule
{Model} 
& {Clean}
& {Noisy}
& {Wild}
& {Expressive}
& {Overall} \\
\midrule
  F5-TTS & -0.82 & -0.48 & -0.48 & -0.95 & -0.68 \\
  CosyVoice~2 & -0.59 & -0.35 & -0.38 & -0.15 & -0.36 \\                                                                                                                                                                                       
  CosyVoice~3 & -0.06 & -0.45 & -0.12 & \underline{0.10} & -0.13 \\                                                                                                                                                                                        
  MaskGCT & \underline{0.15} & \textbf{0.31} & -0.49 & 0.06 & \underline{-0.01} \\                                                                                                                                                                                              
  Qwen3-TTS & 0.08 & -0.46 & -0.38 & \textbf{0.25} & -0.13 \\                                                                                                                                                                                           
  VoxCPM & 0.14 & \underline{0.10} & -0.06 & -0.30 & -0.05 \\                                                                                                                                                                                              
  \midrule                                                                                                                                                                                                                                     
  \RaonModel-0.3B & \textbf{0.54} & -0.24 & \textbf{0.16} & -0.42 & \underline{-0.01} \\                                                                                                                                                                                     
  \RaonModel-1B & 0.00 & 0.00 & \underline{0.00} & 0.00 & \textbf{0.00} \\      
\bottomrule
\end{tabular}
\end{table}

\begin{table}[t]
\centering
\caption{Subjective evaluation results using similarity mean opinion score (SMOS) for perceived speaker similarity to a reference. \textbf{Bold} indicates the best performance, and \underline{underline} indicates the second-best performance.}
\label{tab:smos}
\begin{tabular}{l c c c c c}
\toprule
{Model}
& {Clean}
& {Noisy}
& {Wild}
& {Expressive}
& {Overall} \\
\midrule
  F5-TTS & 3.86 & 3.37 & 3.50 & 3.50 & 3.55 \\                                                                                                                                                                                                 
  CosyVoice~2 & 3.90 & 3.35 & 3.43 & 3.49 & 3.53 \\                                                                                                                                                                                            
  CosyVoice~3 & 3.87 & \textbf{3.63} & 3.46 & 3.40 & 3.58 \\
  MaskGCT & 3.85 & 3.30 & \textbf{3.77} & 3.43 & 3.58 \\                                                                                                                                                                                                
  Qwen3-TTS & 3.89 & 3.47 & 3.55 & 3.48 & 3.59 \\                                                                                                                                                                                              
  VoxCPM & \underline{3.98} & 3.32 & 3.50 & 3.41 & 3.54 \\                                                                                                                                                                                                 
  \midrule                                                                                                                                                                                                                                     
  \RaonModel-0.3B & \textbf{4.01} & 3.55 & 3.39 & \underline{3.54} & \underline{3.60} \\           
  \RaonModel-1B & 3.90 & \underline{3.58} & \underline{3.70} & \textbf{3.64} & \textbf{3.70} \\    
\bottomrule
\end{tabular}
\end{table}

\subsection{Human preference evaluation}
We further conduct human evaluations using both CMOS and SMOS on \RaonEval. In CMOS, listeners compare each baseline against \RaonModel-1B and rate overall naturalness, while in SMOS they assess how similar each synthesized utterance sounds to the reference speech. Further details of the evaluation protocol are described in Appendix~\ref{tab:mturk_setting}. 
As shown in Table~\ref{tab:cmos}, \RaonModel-1B is preferred over the baselines in terms of overall naturalness. Among the closest competitors, MaskGCT records an overall CMOS of -0.01, with gains in the Noisy condition but a notable drop in the Wild condition. 
Table~\ref{tab:smos} reports speaker similarity results. \RaonModel-1B achieves the highest overall SMOS of 3.70, followed by \RaonModel-0.3B at 3.60. 
Across acoustic conditions, \RaonModel-1B maintains strong and stable speaker similarity, achieving the best SMOS in the Expressive condition and the second-best SMOS in the Wild and Noisy condition.
While the 1B model achieves better average objective scores and higher overall speaker similarity, the category-wise subjective results suggest that the 0.3B and 1B variants represent different robustness profiles within a practical scaling range.

\section{Ablation}
\label{sec:ablation}
In this section, we analyze how different aspects of the training data contribute to zero-shot TTS performance. We study two aspects of the training data: (1) the effect of training data source at matched scale and (2) the impact of incorporating large-scale in-the-wild data. All experiments use the Raon-OpenTTS-0.3B model trained for 200K steps and are evaluated on Raon-OpenTTS-Eval, which spans four acoustic conditions~(Clean, Noisy, Wild, and Expressive).
  
  \begin{table}[t]
  \centering                          
  \caption{Comparison of training data source at matched scale (47K hours each). Emilia refers to the English subset used throughout this paper; Pool-Matched-47K is a \RaonPool subset of equal duration. \textbf{Bold} indicates the best performance.} \label{tab:03b-emilia} 
  \resizebox{\columnwidth}{!}{%
  \begin{tabular}{lcccccccccc}                          
  \toprule
  \multirow{2}{*}{Data} &                  
    \multicolumn{2}{c}{Clean} &       
    \multicolumn{2}{c}{Noisy} & 
    \multicolumn{2}{c}{Wild} & 
    \multicolumn{2}{c}{Expressive} &  
    \multicolumn{2}{c}{Overall} \\
  \cmidrule(lr){2-3}\cmidrule(lr){4-5}\cmidrule(lr){6-7}\cmidrule(lr){8-9}\cmidrule(lr){10-11}  
  & WER$\downarrow$ & SIM$\uparrow$ & WER$\downarrow$ & SIM$\uparrow$ & WER$\downarrow$ & SIM$\uparrow$ & WER$\downarrow$ & SIM$\uparrow$ & WER$\downarrow$ & SIM$\uparrow$ \\    
  \midrule
  Emilia          & \textbf{1.28} & 0.593 & 4.02 & 0.641 & 7.35 & 0.470 & 3.60 & 0.505 & 3.33 & 0.558 \\                   
  Pool-Matched-47K   & 1.53 & \textbf{0.631} & \textbf{3.90} & \textbf{0.687} & \textbf{5.97} & \textbf{0.549} & \textbf{3.22} & \textbf{0.546} & \textbf{3.09} & \textbf{0.605} \\ 
  \bottomrule           
  \end{tabular}%
  }            
  \end{table}

 \begin{table}[t]                                                                                                                                                                        
  \centering                                                                                                                                                                              
  \caption{Impact of YouTube-Commons data on zero-shot TTS performance on Raon-OpenTTS-Eval. all data refers to training on \RaonPool (615K hours), while w/o YC denotes the split without \RaonYouTube (280K hours). \textbf{Bold} indicates the best performance.}                    
  \label{tab:youtube-ablation}                                    
  \begin{tabular}{l cc cc cc cc cc}                                                                                                                                                     
  \toprule                                                                                                                                                                                
  Dataset                                                                                              
  & \multicolumn{2}{c}{Clean}                                                                                                                                                    
  & \multicolumn{2}{c}{Noisy}                                                                                                                                                    
  & \multicolumn{2}{c}{Wild}                                                                                                                         
  & \multicolumn{2}{c}{Expressive}                                                                                                                                                      
  & \multicolumn{2}{c}{Overall} \\                                                                                                                                               
  \cmidrule(lr){2-3} \cmidrule(lr){4-5} \cmidrule(lr){6-7} \cmidrule(lr){8-9} \cmidrule(lr){10-11}
  & WER$\downarrow$ & SIM$\uparrow$                                                                                                                                                     
    & WER$\downarrow$ & SIM$\uparrow$                                                                                                                                                     
    & WER$\downarrow$ & SIM$\uparrow$     
    & WER$\downarrow$ & SIM$\uparrow$                                                                                                                                                     
    & WER$\downarrow$ & SIM$\uparrow$ \\                                                                                                                                                  
  \midrule                                                                                                                                                                                
  w/o YC                                                                                                                                                     
    & 2.17 & 0.615                                                
    & \textbf{4.21} & 0.668                                                                                                                                                               
    & 7.62 & 0.523                                                                                                                                                                        
    & 3.43 & 0.540                        
    & 3.73 & 0.590 \\                                                                                                                                                                     
  all data                                                                                                                                                          
    & \textbf{1.72} & \textbf{0.634}      
    & 6.79 & \textbf{0.688}                                                                                                                                                               
    & \textbf{6.15} & \textbf{0.550}                                                                                                                                                      
    & \textbf{2.63} & \textbf{0.560}      
    & \textbf{3.53} & \textbf{0.610} \\                                                                                                                                                   
  \bottomrule                                                                                                                                                                             
  \end{tabular}                           
  \end{table}

\subsection{Training data composition at matched scale}\label{sec:domain-ablation}
To study how \RaonPool compares with a widely used open TTS dataset at matched scale, we compare two models trained on 47K-hour datasets. 
The first model is trained on the English subset of Emilia, a widely used open speech dataset, whereas the second is trained on a \RaonPool subset of equal duration (denoted as Pool-Matched-47K). 
To keep this comparison focused on data source composition, we compare the English subset of Emilia with a subset sampled from \RaonPool rather than \RaonCore; the effect of quality filtering is analyzed separately in Section~\ref{sec:raoncore}.
As shown in Table~\ref{tab:03b-emilia}, Pool-Matched-47K outperforms Emilia on most conditions. Overall WER improves from 3.33\% to 3.09\%, with notable gains in Wild (7.35\%~$\to$~5.97\%) and Expressive (3.60\%~$\to$~3.22\%). Speaker similarity also improves substantially, rising from 0.558 to 0.605. The consistent improvements across acoustic conditions suggest that \RaonPool provides a more effective training signal under a fixed data budget.

\subsection{Incorporating large-scale in-the-wild data}\label{sec:youtube-ablation}
We next analyze the contribution of \RaonYouTube, a large-scale in-the-wild speech dataset included in \RaonPool. To this end, we compare two training configurations: one trained on the full dataset (615K hours), and another trained on a reduced version excluding \RaonYouTube (280K hours).
Because the two configurations differ in both scale and domain composition, this comparison measures their combined effect rather than disentangling the two factors. We intentionally evaluate this combined effect because \RaonYouTube contributes not only additional scale but also real-world acoustic coverage, both of which are part of its practical value as a newly processed TTS training source.
As shown in Table 10, incorporating \RaonYouTube improves performance across several conditions. In particular, WER decreases in Clean (2.17\% → 1.72\%), Wild (7.62\% → 6.15\%), and Expressive (3.43\% → 2.63\%) settings, while speaker similarity also improves across most categories. The largest improvements are observed in the Wild condition, suggesting that in-the-wild data is particularly beneficial for handling highly variable real-world speech.
However, WER degrades in the Noisy condition (4.21\% → 6.79\%), suggesting that the added in-the-wild data improves robustness broadly but does not benefit every acoustic regime equally.

\section{Conclusion}
\label{sec:conclusion}
We presented \textbf{\RaonModel}, an open-data, open-weight zero-shot TTS model with performance comparable to state-of-the-art TTS models trained on proprietary data, together with
\textbf{\RaonPool}, a 615K-hour English speech corpus constructed from diverse sources,
and \textbf{\RaonCore}, a filtered high-quality subset derived from \RaonPool used to train \RaonModel.
By releasing the dataset, filtered subset, and model weights, we support reproducible research and provide a transparent reference for large-scale TTS development.
Experimental results in multiple benchmarks demonstrate that \RaonPool provides sufficient data scale and quality to train TTS models with performance comparable to state-of-the-art systems trained on proprietary data.
We further propose \textbf{\RaonEval}, a robustness-oriented benchmark covering diverse acoustic conditions,
on which \RaonModel also demonstrates strong and consistent performance across a wide range of linguistic complexity and acoustic conditions.

\paragraph{Future work.}
While the current results are encouraging, several directions remain for future improvement.
First, our current study only focuses on English speech data. Extending to multilingual settings can support open-data and open-weight TTS research across a broader range of languages.
Second, \RaonPool aggregates speech from multiple sources with various domains and recording conditions. Developing more effective data mixing or source balancing strategies could further improve model robustness.
Third, \RaonCore is constructed mainly by filtering low-quality samples. Future work could explore ways to recover such data through data correction or speech processing techniques.

\bibliography{reference}

\clearpage
\appendix

\section{Model Architecture Details}
\label{sec:model_config}
Table~\ref{tab:model_configs} summarizes the architectural configurations of the
two model variants: \RaonModel-0.3B and \RaonModel-1B. 
The 0.3B model adopts the original configuration of F5-TTS without modification. 
For the larger variants, we scale up the model capacity by 
increasing the number of transformer layers, attention heads, and embedding 
dimensions while preserving the overall architectural design.

\begin{table}[h]
\centering
\caption{Architecture configurations of \RaonModel models.}
\label{tab:model_configs}
\begin{tabular}{lccc}
\hline
\textbf{Configuration} & \textbf{0.3B} & \textbf{1.0B} \\
\hline
Transformer Layers & 22 & 28 \\
Attention Heads & 16 & 22 \\
Attention Embedding Dimension & 1024 & 1408 \\
Feed-forward Dimension & 2048 & 5632 \\
Text Embedding Dim & 512 & 512 \\
Total Parameters & 336M & 1048M \\
\hline
\end{tabular}
\end{table}

\section{Human Preference Evaluation}
\label{tab:mturk_setting}

We conduct human preference evaluation using two complementary subjective metrics: similarity mean opinion score (SMOS) and comparative mean opinion score (CMOS). These evaluations are designed to measure two distinct aspects of synthesized speech: (1) how similar a generated utterance is to a reference speech prompt, and (2) how its perceptual quality compares to baseline models. As noted in Section~5.4, for each acoustic condition in \textsc{\RaonEval}, we randomly sample 30 evaluation items and collect ratings from annotators recruited through Amazon Mechanical Turk in the United States. Each evaluation item is rated by 6 annotators. For CMOS, the presentation order of the two samples is randomized to avoid position bias. We report the mean score together with the 95\% confidence interval.

\noindent\textbf{SMOS.}
For the SMOS evaluation, annotators are presented with a reference audio sample and a generated audio sample. They are asked to rate how similar the generated speech is to the reference on a 5-point rating scale, with emphasis on speaker similarity, speaking style, acoustic conditions, and background characteristics. More specifically, listeners are instructed to judge whether the generated sample sounds as if it were recorded by the same speaker, in a similar environment, and with a similar speaking style as the reference.

The rating scale is defined as follows:
\begin{itemize}
    \item \textbf{5 (Excellent):} extremely similar to the reference speech
    \item \textbf{4 (Good):} highly similar to the reference speech
    \item \textbf{3 (Fair):} moderately similar to the reference speech
    \item \textbf{2 (Poor):} weak similarity to the reference speech
    \item \textbf{1 (Bad):} very dissimilar to the reference speech
\end{itemize}

\noindent\textbf{CMOS.}
For the CMOS evaluation, annotators are presented with two synthesized speech samples, denoted Audio A and Audio B. They are asked to rate the overall quality of Audio B relative to Audio A using a 7-point comparative scale ranging from $-3$ to $3$. In our evaluation, Audio A and Audio B correspond to outputs from two different models, enabling direct pairwise comparison between our model and each baseline.

Listeners are instructed to assess the overall perceptual quality of Audio B relative to Audio A, taking into account factors such as naturalness, clarity, and comprehensibility. A positive score indicates that Audio B is better than Audio A, a negative score indicates that it is worse, and zero indicates no noticeable quality difference.

The comparative rating scale is defined as follows:
\begin{itemize}
    \item \textbf{3:} much better
    \item \textbf{2:} better
    \item \textbf{1:} slightly better
    \item \textbf{0:} about the same
    \item \textbf{-1:} slightly worse
    \item \textbf{-2:} worse
    \item \textbf{-3:} much worse
\end{itemize}

\noindent\textbf{Evaluation protocol.}
For both SMOS and CMOS, annotators are instructed to use headphones and complete the task in a quiet environment. They are also asked to adjust their listening volume during the training phase and keep it fixed throughout the evaluation. As a quality-control measure, we exclude responses in which an annotator assigns identical scores to all five items on a single evaluation page.

\end{document}